\documentclass[11pt]{article}

\usepackage[version=3]{mhchem} 
\usepackage{amsmath}
\usepackage{gensymb}
\usepackage[colorlinks = true,
linkcolor = blue,
urlcolor = blue,
citecolor = blue,
anchorcolor = blue]{hyperref}
\usepackage{pifont}
\usepackage{ulem}
\usepackage{multirow}
\usepackage{caption}
\usepackage[table,xcdraw]{xcolor}
\usepackage{hhline}
\usepackage{bbold}
\usepackage{setspace} 
\usepackage{ragged2e}
\justifying
\usepackage{etoolbox}
\usepackage{authblk}

\usepackage{graphicx} 
\usepackage{amsmath} 
\usepackage{amssymb} 
\usepackage{hyperref} 
\usepackage{authblk} 
\usepackage{lineno} 
\usepackage{float}
\usepackage{geometry}
\usepackage{jabbrv}

\DefineJournalAbbreviation{Reports}{Rep} 
\DefineJournalAbbreviation{Computational}{Comp}
\DefineJournalAbbreviation{Materials}{Mat}
\DefineJournalAbbreviation{Advances}{Adv}
\DefineJournalAbbreviation{Letters}{Lett}
\DefineJournalAbbreviation{Technology}{Tech}

\usepackage[backend=biber,
    style=phys,
    maxbibnames=5,
    minbibnames=1
]{biblatex}
\AtEveryBibitem{\clearlist{language}}
\DeclareFieldFormat[misc]{title}{#1}

\appto{\bibsetup}{\raggedright}

\title{PtyRANNOSAUR: Ptychography with Robust Artificial Neural Networks Optimized for Sub-Angstrom Accuracy and Ultrafast Reconstruction}

\author[1,2,+]{Kieran~Loehr}
\author[2,3,+]{Rahim~Raja}
\author[1,2]{Xiaochuan~Ding}
\author[2,3]{Jeffrey~Huang}
\author[2,3]{Gillian~Nolan}
\author[2,3]{Sang~hyun~Bae}
\author[1,2,4,]{Bryan~K.~Clark 
\thanks{\href{mailto:bkclark@illinois.edu}{bkclark@illinois.edu}}}
\author[1,2,3,5,]{Pinshane~Y.~Huang \thanks{\href{mailto:pyhuang@illinois.edu}{pyhuang@illinois.edu}}}

\affil[1]{Department of Physics, University of Illinois Urbana-Champaign, Urbana, IL, 61801, USA}
\affil[2]{The Grainger College of Engineering, University of Illinois Urbana-Champaign, Urbana, IL, 61801, USA}
\affil[3]{Department of Materials Science and Engineering, University of Illinois Urbana-Champaign, Urbana, IL, 61801, USA}
\affil[4]{The Anthony J. Leggett Institute for Condensed Matter Theory and IQUIST and NCSA Center for Artificial Intelligence Innovation, University of Illinois Urbana-Champaign, Urbana, IL, 61801, USA}
\affil[5]{Materials Research Laboratory, University of Illinois Urbana-Champaign, Urbana, IL, 61801, USA}
\affil[+]{These authors contributed equally to this work}

\date{}
\begin{document}

\maketitle
\newpage

\begin{abstract} We present PtyRANNOSAUR, a data-driven neural network code that reconstructs atomic resolution electron ptychography data in seconds, 10-100x faster than standard methods. PtyRANNOSAUR uses convolutional autoencoders to map 4D-scanning transmission electron microscopy data to 2D phase images. Each model is trained on a large database of crystal structures and is tailored for a range of experimental parameters, such as accelerating voltage, convergence angle, defocus, and sample thickness. This approach yields high quality reconstructions without requiring any fine-tuning of hyperparameters. In addition, the code handles spatial partial coherence, multiple scattering, and scan position errors, which are critical for state-of-the-art electron ptychography reconstructions. By testing PtyRANNOSAUR on experimental and simulated data, we show that the neural networks accurately reconstruct atomic structures of a broad range of materials systems and can achieve high resolutions of $<0.5$~\AA, comparable to the best iterative reconstructions of the same data. These advances enable near-live, state-of-the-art electron ptychography reconstructions.
\end{abstract}

\section*{INTRODUCTION} 

Electron ptychography is an emerging method that recovers phase images of 2D or 3D objects from 4D scanning transmission electron microscopy (STEM) datasets \cite{Rodenburg2019}. In 4D-STEM, convergent beam electron diffraction (CBED) patterns are acquired using an  angstrom-scale scanning electron beam. By leveraging interference information from overlapping diffraction discs and densely sampled probe positions, electron ptychography recovers the missing phase information needed to form real-space images of the object. Compared to conventional transmission electron microscopy (TEM) and STEM, electron ptychography has several key advantages. Using electron ptychography, deep sub-angstrom resolution ( $<$0.5 \AA) can be obtained in both high-end aberration-corrected electron microscopes and more accessible uncorrected STEMs, unlocking broad access to state-of-the-art microscopy \cite{Nguyen2024Uncorrected}. Electron ptychography can achieve spatial resolutions of 20 pm, beyond conventional optical limits \cite{Nellist1995-ot,Chen2021}, while also having higher dose efficiency than STEM\cite{PENNYCOOK2015160,PennyCook2019DoseEfficiency,Li2025,Chen2020}. Due to these advantages, electron ptychography is rapidly finding broad applications in the characterization of materials and structures at the atomic scale \cite{Miao2025-ye}, including low-dose imaging of soft materials \cite{Kharel2025, Li2025, Li20224d, zhang2023zeolite}, mapping point defects \cite{Jiang2018,Dong2024,BHAT2026114282,chen2024interstitial}, and detecting light elements \cite{Wang2017LaB6,shi2025electron,li2025atomicscaleheterogeneityhydrogenmetal}.

Currently, the utility and widespread adoption of electron ptychography are limited by the computational cost of data processing and the difficulty of achieving reliable, high-quality image reconstructions. State-of-the-art approaches utilize powerful, iterative physics-based solvers \cite{LeePtyRAD2025,Gilgenbach2026-vy,Du2025-wz,Mukherjee2025,Diederichs2024-or} that leverage automatic differentiation to optimize a forward model of electron scattering in a material \cite{Maiden2009,Wakonig2020PtychoShelves,Thibault_2012}. This approach typically requires hundreds to thousands of iterations spanning hours on modern GPUs (e.g. NVIDIA A100). The computation time increases for multislice methods, which are required to model most TEM samples and achieve more accurate reconstructions \cite{Tsai:16,Gao2017-ay,Van_den_Broek2013-hg,Chen2021}. While direct ptychography methods, such as single side band \cite{RODENBURG1993304}, Wigner distribution deconvolution \cite{Rodenburg1992-xc}, or tilt-corrected bright field \cite{Yu2025-hs} operate quickly \cite{liveprocessing2021,fastbinary2020}, they are limited in resolution and do not incorporate incoherence or multiple scattering effects \cite{RODENBURG1993304,MA2026Information}. These challenges severely limit the accessibility and experimental throughput of ptychography and preclude real-time sample feedback during data acquisition with the high image quality and resolution of iterative methods. 

Machine learning (ML) offers a path to dramatically accelerate electron ptychography reconstructions. ML methods have advanced many aspects of electron microscopy \cite{Kalinin2022ReviewML, Kalinin2023-ps,Spurgeon2021-fm}, including acquisition strategies \cite{Schloz2023AdaptiveScanning,Zheng2021-nz,yin2024pear,yin2026autonomous}, denoising \cite{ParkDenoising,Thornley2026-tv,Mohan2022-ll,Mohan2025-ym}, dataset analysis and processing \cite{Lobato2024-mi,WANG2024175,Kong2022-eu,Li2018-fd}, and  image segmentation and classification \cite{Lin2021-ho,ChiaHao2020,Zhu2023seg,Kaur2026-lk,Ortega_Ortiz2025-ml,Groschner2021ML,Rangel_DaCosta2024-kt}. A key advantage of ML for electron ptychography is that data-driven models, such as deep neural networks, excel at learning complex, non-linear relationships \cite{NN_ReviewArticle}, such as those between 4D-STEM data and the 2D phase image. Further, ML can front-load computationally demanding steps to a separate training phase, making individual reconstructions fast and computationally inexpensive, as demonstrated recently for X-ray ptychography \cite{prevML_XRAY, prevML_HOIDN, prevML_newXRAY,hoidn2026singleshotcoherentimagingoverlapfree,PtychoDV}. 

Recent progress in ML for electron ptychography can be categorized into hybrid ``ML + iterative'' and single shot methods.   These methods have so far come with trade-offs between accuracy, generalizability, and speed. While existing single shot ptychography codes \cite{prevML_DeepCDI,prevML_AIRPI} enable rapid reconstructions, they do not account for complex electron scattering physics, variations in the microscope optics, or errors in scan positions;  correspondingly, such methods do not achieve the reconstruction resolution and quality of iterative ptychography. Hybrid approaches aim to achieve the accuracy and broad applicability of iterative methods while accelerating key computational steps using ML \cite{mccray2025deepgenerativepriorsrobust,prevML_Diffusion,prevML_accelerated}, but they are intrinsically slower than single shot ML reconstructions. Electron ptychography reconstruction methods that can operate at or near real-time while also being accurate and generalizable to a broad range of material types, thicknesses, and operating conditions without sacrificing image resolution and quality remain an outstanding need.

\section*{RESULTS AND DISCUSSIONS}

In this work we introduce PtyRANNOSAUR, a 2D atomic resolution electron ptychography reconstruction code that runs in seconds on an NVIDIA A100 GPU for live  reconstructions at or near the speed of data acquisition. By training specialized models across different parameter regimes, PtyRANNOSAUR produces images with quality comparable to, and in some cases exceeding, state-of-the-art iterative ptychography methods across a broad range of materials systems.

\subsection*{Performance on Experimental Data}
\begin{figure*}[htbp]
\centering
\includegraphics[width=1.0\textwidth]{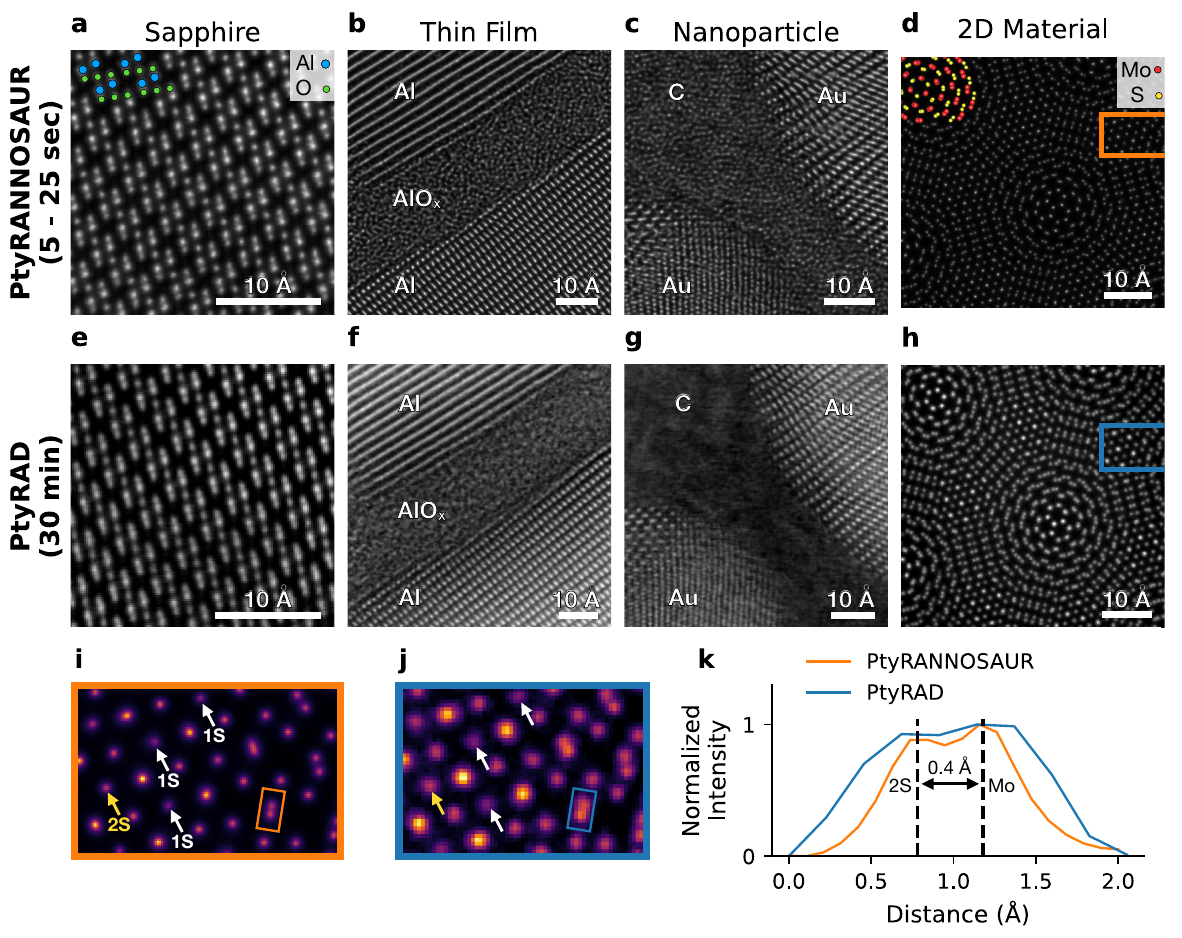}
\caption{Experimental reconstructions using (a-d) PtyRANNOSAUR's neural network approach and (e-h)  iterative methods via PtyRAD.  Data show: single crystalline sapphire, an Al/AlO$_{x}$/Al thin film multilayer, polycrystalline Au nanoparticles on an amorphous carbon support, and twisted bilayer MoS$_2$. For each, PtyRAD reconstructions were run for 30 minutes on an NVIDIA A100 GPU, and single-shot PtyRANNOSAUR reconstructions took 5-25 seconds on the same GPU (5 seconds for the network + 20 seconds for optional position correction). PtyRANNOSAUR reconstructions shown are generated with: (a) Model 20 + rigid stitching, (b) Model 10 + rigid stitching, (c) Model 10 + position correction, (d) Model 2 + position correction (see Methods for additional details). (i,j) Magnified regions of MoS$_2$ with (i) PtyRANNOSAUR  and (j) PtyRAD with arrows labeling 2S and 1S atomic columns. (k) Line profiles averaged across the smaller direction of the indicated rectangles in (i,j) showing resolved projected atomic separations of $<$0.5 \AA.}
\label{fig:Fig1}
\end{figure*}

Figure~\ref{fig:Fig1} compares experimental reconstructions of the same datasets, processed using either PtyRANNOSAUR or PtyRAD, a well-benchmarked and frequently used GPU-accelerated iterative ptychography reconstruction code \cite{LeePtyRAD2025}. Iterative reconstructions were performed using standard PtyRAD hyperparameters (see Methods). Each PtyRANNOSAUR reconstruction is produced in a single forward pass in approximately 5 seconds on a single NVIDIA A100 GPU, whereas the PtyRAD reconstructions were obtained at 30 minutes. Four specimens are shown:  a) bulk sapphire [100], b) a thin film multilayer of Al/AlO$_{x}$/Al, c) gold nanoparticles on an amorphous carbon support, and d) twisted bilayer MoS$_2$. These samples span a range of thicknesses ($\approx 1 - 25$ nm), material types, and geometries typical of TEM specimens. 

Across all samples, PtyRANNOSAUR accurately reconstructs atomic resolution images and produces outputs comparable to those obtained via conventional iterative methods.  For example, the reconstruction of sapphire ($\alpha$-Al$_2$O$_3$) in Fig.~\ref{fig:Fig1}a shows the expected atomic structure, as indicated in the overlaid atomic structure, with heavier Al columns appearing brighter than the lighter O columns, and obtaining strong agreement with the  iterative PtyRAD result in Fig.~\ref{fig:Fig1}e.  PtyRANNOSAUR also successfully generalizes to more complex structures. The thin film multilayer and nanoparticles in Fig.~\ref{fig:Fig1}b-c  are aperiodic, containing multiple distinct crystals separated by amorphous regions. Again, image reconstructions from PtyRANNOSAUR obtain similar structures to PtyRAD. These data show that PtyRANNOSAUR is able to successfully reconstruct crystalline structures with defects, interfaces, and crystals tilted off major zone-axes. 

Fig.~\ref{fig:Fig1}d tests PtyRANNOSAUR's performance on a moiré structure of twisted bilayer MoS$_2$. Twisted 2D materials are valuable test specimens for electron microscopy because they provide a range of 2D projected spacings between atoms, enabling direct measurements of spatial resolution. Fig.~\ref{fig:Fig1}i,j show magnified regions of the twisted \ce{MoS2} lattice. In these regions, atomic columns appear sharper and better resolved in the neural network reconstruction. For example, the orange and blue boxes indicate the same two atomic columns in reconstructions from PtyRANNOSAUR and PtyRAD, with corresponding line profiles in Fig.~\ref{fig:Fig1}k. The two atomic columns with peaks spaced $0.4$~\AA\;apart are barely separated in the PtyRAD image but clearly separated in the PtyRANNOSAUR reconstruction, indicating our network's ability to achieve deep sub-angstrom resolution  of $<0.5$ \AA.  PtyRANNOSAUR  also detects individual vacancies. In Fig.~\ref{fig:Fig1}i,j, the white arrows indicate single sulfur vacancies, which have reduced intensity compared to a column of two sulfur atoms in projection indicated by the yellow arrows.  These data demonstrate the neural network's high sensitivity to defect structures and its ability to locate single atom defects within materials, without requiring material-specific training.

\subsection*{Data Simulation and Model Training}

\begin{figure*}[htbp]
\centering
\includegraphics[width=1.0\textwidth]{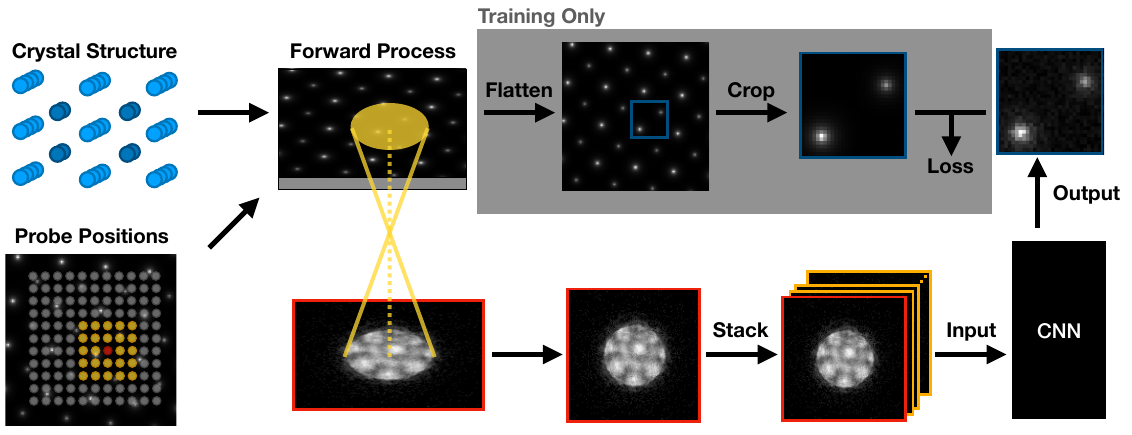}
\caption{Schematic for training and generating ptychography reconstructions of local patches of objects using PtyRANNOSAUR. 
An electron scattering forward process obtains diffraction patterns from a given crystal structure at specified probe positions. A $5 \times 5$ grid (red, gold dots) of 25 diffraction patterns form the input to the convolutional neural network (CNN), which outputs a 2D phase image centered at the central probe position (red dot). During inference, the forward process is experimental 4D-STEM data acquisition and the output is the ptychography reconstruction (which produces patches that are stitched to form the full object - see Fig.~\ref{fig:Fig5}). In training, atomic potentials and 4D-STEM data are simulated with abTEM and then flattened and cropped (``Training Only" path) to create input-target pairs using crystal structures from the Materials Project. The structural similarity index measure (SSIM) between the cropped patch and the CNN output is used as a loss function to train the network.
}
\label{fig:Fig2}
\end{figure*}

To train a PtyRANNOSAUR model, we generate a large database of 64 million pairs of input potentials and diffraction pattern sets by pulling 16,000 stable, experimentally observed materials from the Materials Project \cite{MaterialsProject}. 3 \AA\ $\times$ 3 \AA\ projected phase images are paired with corresponding sets of simulated CBED patterns from a of $5 \times 5$ grid. The projected phase images serve as model targets while the CBED patterns are used as model inputs. To reconstruct larger fields of view, overlapping local patches are predicted independently and stitched together to form the final image.

The CBED patterns are generated using multislice electron scattering simulations in abTEM \cite{10.12688/openreseurope.13015.2}. For each model, 4D-STEM diffraction patterns are simulated using a combination of fixed microscope parameters, including accelerating voltage, convergence angle, and scan step size, together with variable parameters such as sample thickness and defocus. Introducing variability in these parameters increases the robustness of the learned inverse mapping and reduces sensitivity to experimental uncertainty. 

Our methods differ from previous ML-based ptychography in two key ways. First, unlike models trained on stock internet images \cite{prevML_DeepCDI}, our neural network can implicitly learn structural characteristics relevant to realistic crystalline materials (also used recently in Ref.~\cite{prevML_AIRPI}). Second, our method explicitly incorporates multiple scattering through the multislice forward model used to generate the training data. This enables improved reconstruction accuracy and allows the network to generalize to thicker samples.

We present three models, named according to the average sample thickness used during training. Model 2 is designed for ultra-thin samples such as 2D materials: it was trained with an 80 kV accelerating voltage and materials thicknesses of 2 $\pm$ 0.6 nm. Models 10 and 20 were trained to accommodate standard TEM lamella, using a 300 kV accelerating voltage and nominal sample thicknesses of 10 $\pm$ 2 nm and 20 $\pm$ 2 nm. Complete training values are summarized in Table~\ref{table:modelparams}.

\begin{table}[h!]
    \centering
    \begin{tabular}{l c c c}
    \hline
    \textbf{Model} & \textbf{Model 2} & \textbf{Model 10} & \textbf{Model 20} \\
    \hline
    Thickness (nm) & $2\pm0.6$ & $10\pm2$ & $20\pm2$ \\
    Accelerating voltage (kV) & 80 & 300 & 300 \\
    Convergence angle (mrad) & 25 & 25 & 25 \\
    Defocus (nm) & $7.5\pm2$ & $5.0\pm2$ & $5.0\pm2$ \\
    Reciprocal space sampling (mrad/px) & 1.67 & 0.86 & 0.86 \\
    Scan step size (\AA) & 0.43 & 0.5 & 0.5 \\
    Source size blur (\AA) & $0.35\pm0.15$ & $0.35\pm0.15$ & $0.35\pm0.15$ \\
    Detector size (pixels)& 128& 128&128\\
    \hline
    \end{tabular}
    \caption{Training parameters used to generate the three models presented in this manuscript. Defocus is defined as the distance of the crossover above the top of the sample.}
    \label{table:modelparams}
\end{table}

For training, we define a loss function based on the structural similarity index measure (SSIM), which quantifies similarities in luminance, contrast, and structure between two images\cite{2004SSIM}. The loss is defined as $\mathcal{L} = 1-\text{SSIM}$ between the network output and the ground truth local phase image, such that a perfect reconstruction yields an SSIM of 1 and a loss of 0. Each model in this work was trained for roughly 64 hours on a single NVIDIA A100 GPU. Importantly, once trained for a specific parameter set, the model can be used indefinitely for rapid reconstructions.

\subsection*{Performance on Simulated Data}

\begin{figure*}[htbp]
\centering
\includegraphics[width=1.0\textwidth]{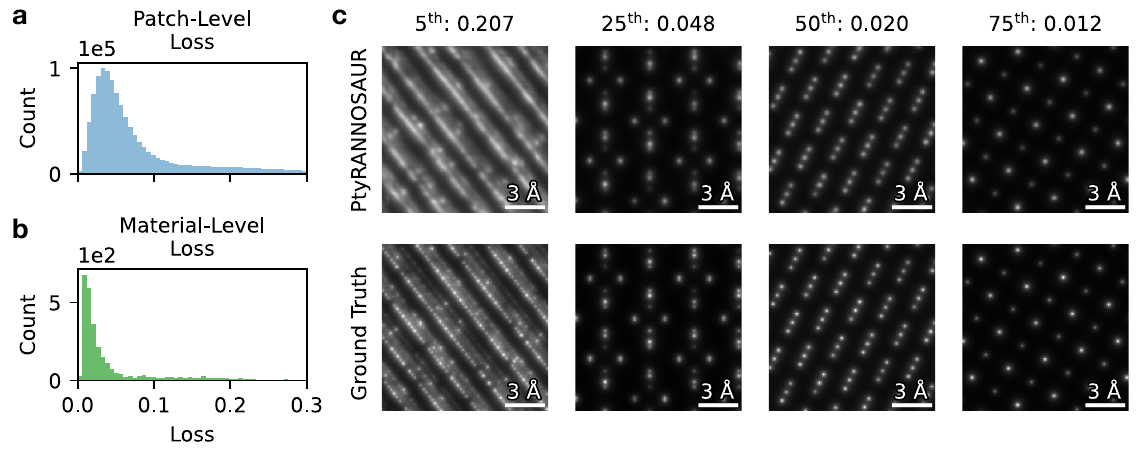}
\caption{Neural network reconstructions from Model 10 on simulated validation materials. 
(a,b) SSIM loss distributions evaluated on 2,950 validation materials. (a) Patch-level loss across all individual network outputs. (b) Material-level loss after stitching patches to form the full image. (c) Representative reconstructions, labeled by the percentile and material-level loss, show (from left to right) TeOF$_2$, NaBF$_4$, TiNiAs, and Sr$_3$Fe$_2$Cu$_2$Se$_2$O$_5$.}
\label{fig:Fig3}
\end{figure*}

In Figure \ref{fig:Fig3}, we assess the performance of PtyRANNOSAUR. We evaluate reconstructions on simulated 4D-STEM datasets from a validation set of 2,950 materials excluded from the training set, and simulated with identical parameters. We quantify reconstruction accuracy using the same SSIM loss, $\mathcal{L} = 1-\text{SSIM}$, used in training. The patch-level loss distribution of Model 10 (Fig.~\ref{fig:Fig3}a) shows a median loss of 0.053 with a long decaying tail. After stitching overlapping patches into full reconstructions, the median material-level loss drops to 0.020 (Fig.~\ref{fig:Fig3}b), consistent with expected error reductions from patch averaging.

To visualize the reconstructed image quality, Fig.~\ref{fig:Fig3}c shows representative reconstructions at the 5th (poorest), 25th, 50th, and 75th (best) percentiles of the material-level loss alongside their corresponding ground truth phase image. At the 75th and 50th percentiles, the network faithfully reproduces periodic lattice structures, capturing both atomic positions and qualitative  relative intensities. Generally, we find that higher loss is observed in  materials with very closely spaced atomic columns. For example, in the 25th percentile image, columns separated by 0.4~\AA{} are mostly resolved, while the columns separated by 0.2~\AA{} are not, elevating the loss. The highest losses are obtained for structures away from a major zone axis and of increased complexity, such as the  5th percentile image in Fig.~\ref{fig:Fig3}c.  Here, the network recovers only the coarse  structure, consistent with the close projected spacing and added difficulty of the tilted crystal orientation. Additional examples of high loss reconstructions are shown in Fig.~\ref{fig:FigS4}.

Broadly, higher losses occur in regimes that are intrinsically difficult for ptychography reconstructions. This suggests that the network has learned a physically meaningful representation of the inverse problem rather than overfitting to the training distribution. These results demonstrate that PtyRANNOSAUR generalizes effectively to unseen materials and produces accurate reconstructions across a broad range of crystal structures. 

\subsection*{Robustness to Parameter Variations}

\begin{figure*}[htbp]
\centering
\includegraphics[width=1.0\textwidth]{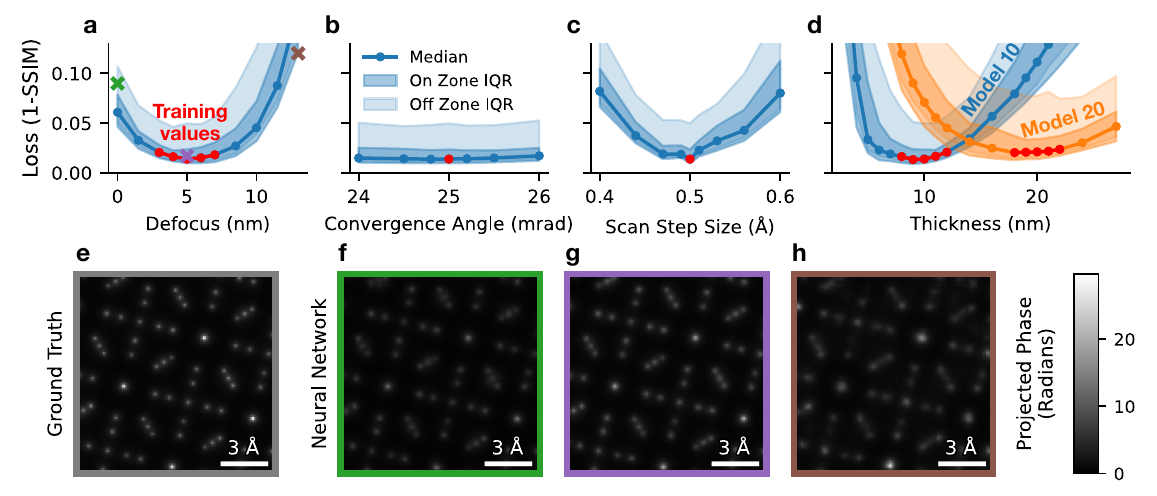}
\caption{PtyRANNOSAUR robust performance with parameter variations. (a-d) Reconstruction loss  for 200 validation materials, varying either defocus, convergence angle, scan step size, or sample thickness. For each parameter, the median loss and interquartile range (IQR) are shown for 140 validation materials that are simulated along zone axes. Additionally, the IQR including the 60 validation materials simulated at a slight tilt off zone axes is shown. The training values are highlighted in red. Model 10 is used in (a-c) while both Model 10 (blue) and Model 20 (orange) are plotted in (d) to show how different models can be trained for different thicknesses. (e) Ground truth atomic potentials of an example validation material, Lu$_6$Mn$_{23}$. (f-h) Corresponding Model 10 reconstructions of Lu$_6$Mn$_{23}$ with simulated defocus values of 0, 5, and 13 nm respectively. Colored x markers in (a) show the loss associated with these reconstructions.}
\label{fig:Fig4}
\end{figure*}

In Figure~\ref{fig:Fig4}, we assess PtyRANNOSAUR's ability to generalize across changes in defocus, convergence angle, scan step size and sample thickness. Fig.~\ref{fig:Fig4}a–c report the median reconstruction loss and interquartile range (IQR) as each parameter is varied around its training value for simulated 4D-STEM datasets from 140 on zone validation materials. The IQR is also shown with the addition of 60 off zone validation materials. Overall, the network demonstrates strong robustness to parameter variations typical of experimental 4D-STEM acquisition. For example, by heuristically defining a cutoff of successful reconstruction at $\mathcal{L} = 0.05$ by inspecting losses and their corresponding qualitative image quality, we find that Model 10 achieves high-quality reconstructions for defocus values in an approximately $8$ nm range around the mean training value (Fig.~\ref{fig:Fig4}a). For the same model, we obtain tolerance ranges of $0.12$ \AA~for scan step size, $10$ nm for thickness, and at least $2$ mrad for convergence angle (Fig.~\ref{fig:Fig4}b-d).

The results show that the network maintains high reconstruction accuracy across a broad range of experimental conditions and reliably handles realistic fluctuations in acquisition parameters. These ranges are similar to tolerance ranges for iterative ptychography. For example, PtyRANNOSAUR's defocus tolerance range of $8$ nm is comparable to the $\pm 5$ nm range quoted for PtyRAD \cite{LeePtyRAD2025}. 

As shown in Fig.~\ref{fig:Fig4}e-h, outside these regimes the reconstruction accuracy decreases, though the overall structure can still be captured. For example, in Fig.~\ref{fig:Fig4}f, with a loss of $0.09$, the closest projected atomic distance of $0.55$~\AA{} is still resolved, but the image exhibits reduced intensity overall. Fig.~\ref{fig:Fig4}h, with a loss of $0.12$ shows the correct overall structure but no longer resolves these close atomic columns.

Training multiple models further extends the accessible parameter space beyond the tolerance of any individual model. Fig.~\ref{fig:Fig4}d compares the performance of Models 10 and 20 for varying sample thicknesses. Their combined coverage enables accurate reconstructions over 5-27 nm, approaching typical upper thickness limits accessible to iterative ptychographic reconstruction methods without energy filtering \cite{Yang2025ThickPtycho}. In practice, users could run multiple models in parallel to span wide experimental regimes without a significant decrease in speed.

In contrast to iterative solvers, PtyRANNOSAUR removes the need for explicitly tuning reconstruction parameters by leveraging models trained at approximate known operating conditions and which generalize sufficiently well to enable accurate experimental reconstructions in a single pass. Reconstruction quality varies with the precise initialization of parameters including microscope settings (e.g., defocus and convergence angle), sample properties (e.g., thickness), and algorithmic hyperparameters (e.g., group size and regularization), many of which can vary between experimental sessions and are unknown or only approximately known  \cite{Wakonig2020PtychoShelves, LeePtyRAD2025}. Achieving optimal reconstructions typically requires multiple runs to tune these parameters, compounding the computational costs of ptychography reconstruction and requiring manual tuning or automated approaches such as Bayesian optimization \cite{Cao2022Bayesian}. 

\subsection*{Stitching Algorithms}

\begin{figure*}[htbp]
\centering
\includegraphics[width=0.9\textwidth]{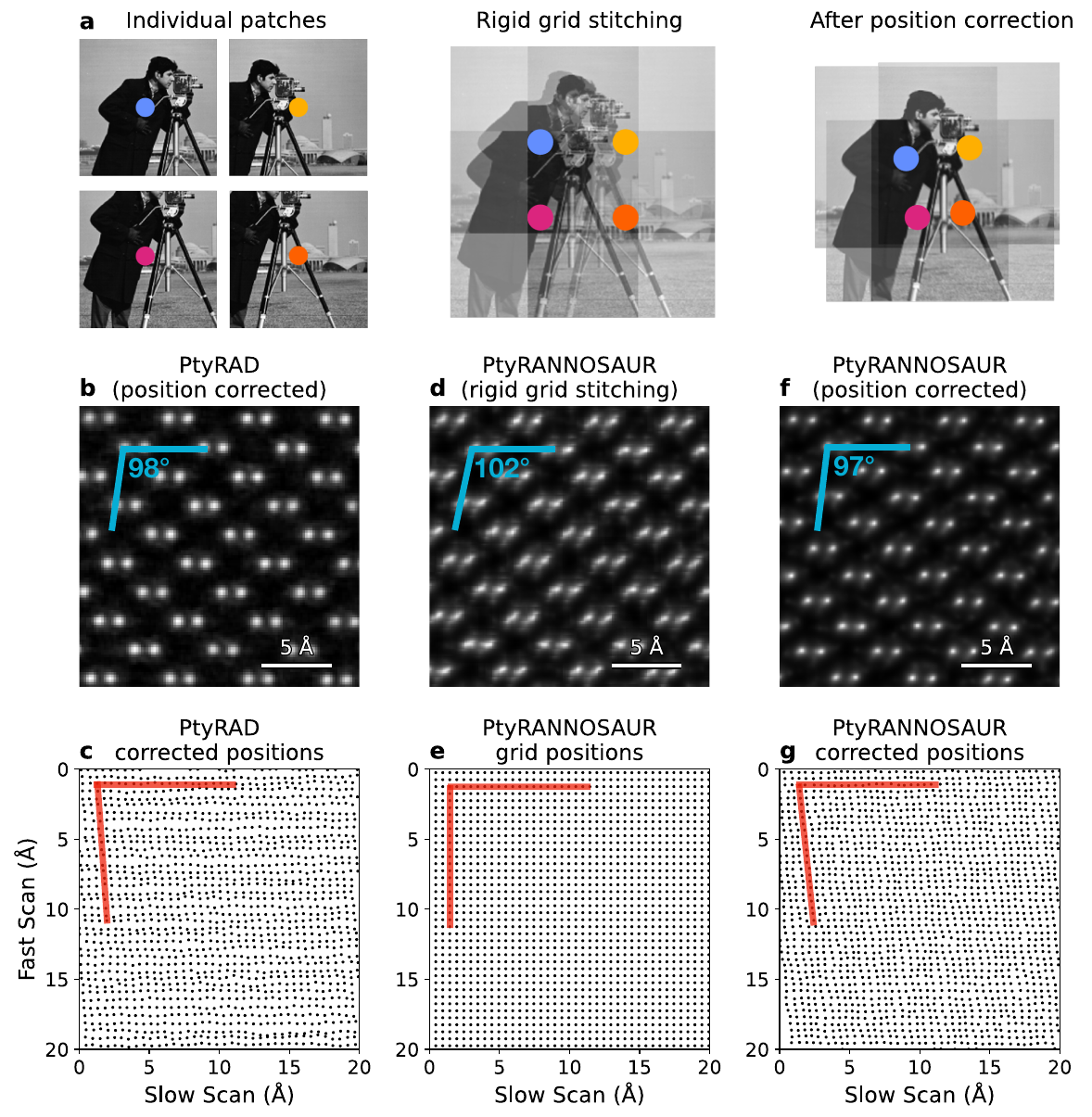}
\caption{Stitching algorithms for combining PtyRANNOSAUR output patches into a full phase image.  (a) Schematic approach for stitching four adjacent patches (left) using either rigid grid positions (center) or position correction via cross correlation (right). (b,d,f) Ptychographic reconstructions of an experimental 4D-STEM dataset of  $\approx$ 20 nm thick silicon (110) acquired under stage drift. Reconstructions were generated using either (b) iterative ptychography run for 5000 iterations over 20 hours, or (d,f) PtyRANNOSAUR using Model 20 with (d) rigid grid stitching or (f) cross correlation-based position correction. Blue annotations illustrate measured angles between the (001) and (1$\overline{1}$0) planes, which should be 90$\degree$. (c,e,g) Corresponding positions for each method.  Dots represent probe positions (in c) or positions used for patch stitching (in e,g). Red lines indicate shear corrected by probe/patch positions.}
\label{fig:Fig5}
\end{figure*}

Our ML architecture outputs angstrom-scale, overlapping patches which need to be assembled to form the full object, as shown in Figure~\ref{fig:Fig5}a.  Rigid grid stitching performs well on experimental datasets with minimal drift, but can create blurring or shear in experimental datasets that contain stage drift, time-varying electromagnetic fields, or jitter in the probe positions \cite{MAIDEN201264}. 
To address this, we implement an optional position correction module that runs in approximately 20 seconds on a typical CPU. Fig.~\ref{fig:Fig5}b,d,f show stitched results for Si [110] acquired under a large drift of $\approx$ 1.5 nm/minute, as measured using ptychography position correction. As shown in Fig.~\ref{fig:Fig5}f, PtyRANNOSAUR's position-corrected reconstruction is visibly sharper than the rigid-grid result (Fig.~\ref{fig:Fig5}d) and produces comparable results to PtyRAD (Fig.~\ref{fig:Fig5}b) using its built-in position correction \cite{LeePtyRAD2025}. Position correction also reduces shear artifacts in the image, making the angle between columns closer to the true value of 90 degrees, though full shear correction is not obtained.  These improvements are achieved even though position errors are not included during training, suggesting that patch consistency provides sufficient structure for partial correction.

\subsection*{Uses and Limitations}

There are three steps to using PtyRANNOSAUR: generating simulated data, training the model, and inference (or image reconstruction). To make PtyRANNOSAUR broadly accessible, the trained models used in this work are available through a user-friendly Jupyter notebook interface. Through this interface, users can perform reconstructions using our existing models  on platforms such as Google Colab, bypassing the need to simulate data and train a new model.  Users may also train custom models for their acquisition conditions. Simulating training data requires nearly 500~GB of storage space and takes approximately 700 CPU hours, while training a PtyRANNOSAUR model takes approximately 64 NVIDIA A100 GPU hours. Once this process is completed, each model produces reconstructions in 5 seconds on a NVIDIA A100 GPU.

While PtyRANNOSAUR is robust to moderate deviations in acquisition parameters, its performance ultimately depends on the similarity between the input data and the training distribution. When applied to data outside its training regime, PtyRANNOSAUR can produce systematic errors in the reconstruction such as intensity errors,  reduced contrast, or blurring between atomic columns (see Supplementary Fig.~\ref{fig:FigS3}). As such, it is critical to select an appropriate model and cross-validate results with other methods.

These challenges are not unique to PtyRANNOSAUR; similarly, in iterative ptychography, reconstruction failures often reflect limitations in the underlying experimental data. In this sense, poor outputs from PtyRANNOSAUR should not always be interpreted as model failure, but rather as a rapid diagnostic of data quality issues—such as miscalibrated defocus, damage, contamination, or insufficient sampling—that would also degrade iterative reconstructions. Generally, we find that in-range data that reconstruct poorly with PtyRANNOSAUR also fail to converge with iterative methods.  A key advantage of our neural network approach is that it can surface data issues much more quickly than conventional iterative methods, allowing the user to quickly optimize data acquisition. 

We envision PtyRANNOSAUR's applications in three main capacities:  to generate priors to speed iterative reconstructions, to rapidly screen samples and data, or as an alternative to iterative ptychography for obtaining 2D phase images. As a prior, PtyRANNOSAUR outputs could be used as initial guess objects or to rapidly fine-tune reconstruction hyperparameters to reduce convergence times.  Additionally, because PtyRANNOSAUR operates faster than typical 4D-STEM acquisitions, it can provide near-live feedback so users can locate features of interest and optimize data acquisition during instrument operation, rather than relying on next-day ptychography reconstructions.   In cases where 2D images are sufficient, PtyRANNOSAUR models can be used to replace iterative ptychography, enabling robust, high-throughput, high resolution reconstruction without parameter tuning.  These capabilities are particularly important for techniques where large numbers of datasets must be reconstructed at high quality, such as in-situ \cite{HaoInSitu2025} experiments or combinations of tomography and ptychography \cite{Chang2020-cj,Dong2025-lu,Ding2022-pp}.

In conclusion, our neural network–based ptychography framework provides a practical route for  fast, robust reconstructions that can be readily integrated into experimental workflows. By significantly reducing the computational cost and demonstrating resilience to parameter variations, PtyRANNOSAUR allows near-live reconstruction on materials up to 25 nm in thickness. This approach can be tailored for individual microscopes and data acquisition pipelines, allowing experimentalists to create personalized neural networks tuned for their systems and experimental parameters. More broadly, our methods demonstrate the capability to obtain electron ptychography reconstructions within seconds, newly enabling accessible, near-live imaging with deep sub-angstrom resolution.

\section*{METHODS} 
\subsection*{Experimental Datasets}
Fig.~\ref{fig:Fig1} experimental datasets were all acquired using a Thermo Fisher Scientific Themis Z aberration-corrected STEM, with acquisition parameters outlined in  Table \ref{table:modelparams}.  We fabricated the twisted bilayer MoS$_2$ sample using standard exfoliation procedures \cite{Yichao2025Phasons}. The sapphire, silicon, and aluminum trilayer junction cross-section samples cross-sectional TEM samples were fabricated using standard lift-out procedures using a Ga+ focused ion beam instrument (FEI Helios 600i Dual Beam FIB-SEM), with protective layers of amorphous carbon and platinum to minimize damage to the sample surface. A final milling voltage of 2 kV was used to reduce surface damage, and a cryo-can was used to minimize redeposition. 

In PtyRAD, we use a single slice reconstruction for the twisted bilayer MoS$_2$ dataset. For all non-2D samples, we use multislice iterative ptychography. We used the samples scripts provided by PtyRAD and using our experimental parameters. All other parameters were unchanged. We use the computed thickness estimate from iterative ptychography throughout this work. 

\subsection*{Training Data Simulation}
Material structures using in training were obtained by filtering structures from Materials Project \cite{MaterialsProject} that were stable and have unit cell lengths under 16~\AA. Each structure was oriented perpendicular to the first and second lattice vector in the unit cell, which may not correspond to a zone axis for monoclinic and triclinic materials. These structures were then tiled to generate a sample of desired dimensions -- a z-thickness matching the target sample thickness and a lateral x-y range large enough to avoid edge effects and achieve a minimum sampling in reciprocal space. The structures are modified to include random vacancies, substitutions, and twists, as detailed in Supplementary Table~\ref{table:supp_defects}, to prevent the model from overfitting to perfect crystals. The 2D projected potential is calculated from atomic structures using an independent atom model \cite{10.12688/openreseurope.13015.2}. The generated potential is proportional to the phase of the transmission function. Ptychography algorithms reconstruct this phase as the phase image. \cite{Kirkland2016ComputationIE}

The multislice forward process is simulated with a slice thickness of 2~\AA{} \cite{Kirkland2016ComputationIE}.
We do not include frozen phonons in the simulations. Due to the finite size of the electron source and mechanical/electrical instabilities, TEM instruments contain partial spatial coherence \cite{Nellist1995-ot}.  This reduces phase contrast in the CBED patterns by effectively averaging over neighboring probe positions. In iterative ptychography, this effect is modeled using mixed-state probe eigenmodes \cite{Odstrcil:16,THIBAULT2009338,Thibault2013-an}. For model training, we use an abTEM function that incorporates partial spatial incoherence by convolving the 4D-STEM data with a Gaussian kernel across the scan dimensions, where the standard deviation corresponds to the estimated instrument source size of  $0.35 \pm 0.15$~\AA{} \cite{Nguyen2024Uncorrected, 10.12688/openreseurope.13015.2}. For parameters sampled over a range of values, including partial spatial incoherence, defocus, and material thickness, values were drawn uniformly from the specified range for each material.

To construct data for PtyRANNOSAUR, we created a large database of 6.4 million training pairs and 1.2 million validation pairs. These come from 18,950 materials, each containing 400 input-target pairs taken from a small, roughly $1$ nm $\times$ $1$ nm region. Each pair consists of a target, the 2D projected phase image for a $3$~\AA{} $\times$ $3$~\AA{} region of a crystalline material, and an input, a set of 25 simulated CBED patterns from this region. The small size of the outputs prevents network bias towards perfectly periodic structures. To improve generalization, we augment the dataset by applying valid symmetry operations, consisting of rotations and reflections, to both the input CBED patterns and corresponding target phase images, increasing the effective size of the training dataset by a factor of eight.

\subsection*{Network Architecture}
The network architecture shown in Supplementary Fig.~\ref{fig:FigS1} is a fully convolutional autoencoder designed to learn the inverse mapping from measured ptychography data to the the real-space projected phase image \cite{AutoencoderReview}. The encoder (down-convolution blocks) progressively reduces the spatial resolution of the input while increasing the number of feature channels. In deeper layers, the network has access to increasingly nonlocal correlations, which are essential in ptychography where each measurement contains overlapping information about the sample. At the bottleneck, the latent convolution block provides a compressed, feature-rich representation of the original input. The decoder (up-convolution blocks) then reconstructs the spatial structure by progressively increasing resolution. Because the reconstruction problem is ill-posed, the upsampling pathway is critical for recovering fine features of atomic data learned from the training data. The hyperparameters of the autoencoder architecture used in all models are shown in Supplementary Table~\ref{table:autoencoderhyperparams}.

\subsection*{Network Optimization}
All neural network implementations were done with JAX \cite{jax}. The loss curves shown in Supplementary Fig.~\ref{fig:FigS2} correspond to the final models used in this work and are representative of the typical optimization trajectory. Training was performed using the Adam optimizer with a learning rate of  $1 \times 10^{-3}$ and a batch size of 32. Under these conditions, all models exhibit smooth and steady decreases in loss, indicating that the optimization landscape is well-behaved for this problem. We do not observe significant plateaus or oscillations, and find that this optimization is repeatable with different initializations of the network and training data.

\subsection*{Stitching Algorithm}
During training, the simulated data is scanned on an exact grid, so we utilize rigid grid stitching, in which patches are averaged at their nominal grid locations. Over 40 patches are summed for each  pixel of the final image. For experimental data, we correct for translational misalignment between patches using phase cross-correlation \cite{kuglin1975phase}, which estimates sub-pixel shifts between overlapping regions. These shifts are then used to place each patch onto a common reconstruction canvas. The corrected reconstruction is obtained by globally combining all patches with interpolation and normalization by the local patch density, ensuring consistent weighting across the field of view. This effectively enforces a self-consistent geometry across overlapping patches. 

To estimate the displacement vectors between neighboring patches, a Hamming window is applied to both patches, and the phase cross correlation between the patches is computed following the implementation in scikit-image \cite{scikit-image}.  
Because the individual patches contain relatively few atoms, there are often spurious peaks in the cross-correlation where one atom is incorrectly correlated with a different atom. Therefore, the peaks within the cross-correlation are multiplied by a penalty factor based on how far the peak is from the expected scan position shift, after which the peak with the largest value is selected. Sub-pixel displacement estimates are obtained by locally up-sampling the cross-correlation \cite{Guizar-Sicairos2008,scikit-image}. Displacement vectors that still deviate excessively from the expected scan position shift are discarded. Then, we optimize the patch positions globally using a cost function which uses the displacement vectors found by cross correlation and the expected scan shift. 

The position correction module is intended as a lightweight post-processing step to improve sharpness by exploiting consistency between overlapping neural network patches. Like any position correction code, it is not guaranteed to improve all datasets, and may misinterpret structural variation as drift and introduce artificial shear. Users should validate corrected reconstructions carefully against known symmetries and expected crystal geometry. Previous work on position correction algorithms \cite{Du:24} showed that iterative reconstructions initialized with corrected positions predicted patches accelerated convergence and reduced position error for iterative ptychography. PtyRANNOSAUR may offer a similar benefit for electron ptychography datasets. 


\section*{DATA AVAILABILITY}
The data sets are available on Zenodo: \url{https://doi.org/10.5281/zenodo.20836142}.

\section*{CODE AVAILABILITY}
The codes are available on github: \url{https://github.com/ClarkResearchGroup/ptyrannosaur}.

Jupyter notebook: \url{https://clarkresearchgroup.github.io/ptyrannosaur/ptyrannosaur.html}.

\section*{ACKNOWLEDGEMENT}
We thank Dr. Chia-Hao Lee for assistance with the PtyRAD code. We thank Zachary Martin and Katie Cauffiel for testing PtyRANNOSAUR. 
This research was carried out in the Materials Research Laboratory Central Facilities, University of Illinois.
This work was supported by AFOSR grant number AF FA9550-23-1-0690. The authors used facilities and instrumentation supported by NSF through the I-MRSEC (DMR-2309037). 
This work made use of the Illinois Campus Cluster, a computing resource that is operated by the Illinois Campus Cluster Program (ICCP) in conjunction with the National Center for Supercomputing Applications (NCSA) and which is supported by funds from the University of Illinois Urbana-Champaign. 
This research was supported in part by the Illinois Computes project which is supported by the University of Illinois Urbana-Champaign.
This work used the Delta system at the National Center for Supercomputing Applications [award OAC 2005572] through allocations MAT240033 and MAT240102 from the Advanced Cyberinfrastructure Coordination Ecosystem: Services and Support (ACCESS) program, which is supported by National Science Foundation grants \#2138259, \#2138286, \#2138307, \#2137603, and \#2138296.

\section*{AUTHOR CONTRIBUTIONS}
K.L. (Conceptualization, Data simulation, Formal analysis, Investigation, Methodology, Software, Visualization, Validation, Writing - original draft)
R.R. (Conceptualization, Data simulation, Data acquisition, Formal analysis, Investigation, Methodology, Visualization, Validation, Writing - original draft)
X.D. (Conceptualization, Investigation, Methodology)
J.H. (Validation, Software - position correction)
G.N. (Data acquisition - Silicon)
S.B. (Data acquisition - MoS$_2$)
B.K.C. (Conceptualization, Supervision, Writing - review)
P.Y.H. (Conceptualization, Supervision, Writing - review)

\section*{COMPETING INTERESTS}
The authors declare no competing financial interest.

\section*{ADDITIONAL INFORMATION}
\noindent\textbf{Correspondence} and requests for materials should be addressed to Pinshane Y. Huang or Bryan K. Clark.

\clearpage
\appendix
\section*{Supplementary Information}

\setcounter{figure}{0}
\renewcommand{\thefigure}{S\arabic{figure}}

\setcounter{table}{0}
\renewcommand{\thetable}{S\arabic{table}}

\setcounter{equation}{0}
\renewcommand{\theequation}{S\arabic{equation}}

\begin{figure}[h!]
\centering
\includegraphics[width=1.0\textwidth]{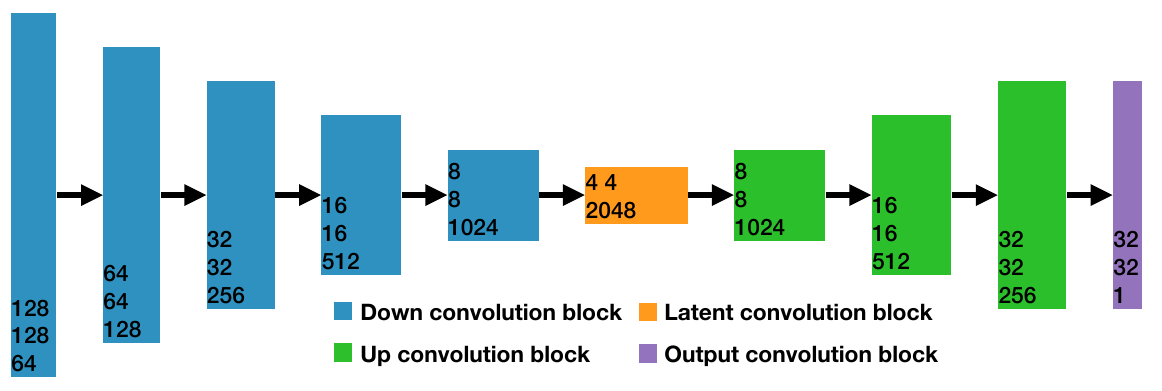}
\caption{Schematic of the convolutional autoencoder architecture used for ptychography reconstruction. Each block consists of two units of a convolutional layer, followed by a batch normalization layer, and a leaky ReLu activation layer. The network consists of a sequence of down-convolution blocks (yellow) that progressively reduce spatial resolution while increasing feature depth with a pooling layer at the end, followed by a latent convolution block (green), and a sequence of up-convolution blocks (blue) that restore spatial resolution by upsampling before each layer. The final output convolution block (purple) maps features to the reconstructed image. The output convolution block does not include batch normalization and replaces the final leaky ReLU activation function with a normal ReLU to ensure the potential is always positive. For each block, the first two numbers denote the spatial dimensions of the convolutional layer, and the third number denotes the number of filters.}
\label{fig:FigS1}
\end{figure}

\begin{table}[h!]
    \centering
    \begin{tabular}{lc}
    \hline
    \textbf{Hyperparameter} & \textbf{Value} \\
    \hline
    Number of down-convolution blocks & 5 \\
    Number of up-convolution blocks & 3 \\
    Base number of filters & 64 \\
    Convolution kernel size & 3 \\
    Pooling/upsampling size & 2 \\
    BatchNorm Momentum & 0.5 \\
    Leaky ReLU slope & 0.01 \\
    Stride & 1 \\
    \hline
    \end{tabular}
    \caption{Neural network hyperparameters used for all PtyRANNOSAUR Models.}
    \label{table:autoencoderhyperparams}
\end{table}

\begin{figure}[h!]
\centering
\includegraphics[width=0.95\textwidth]{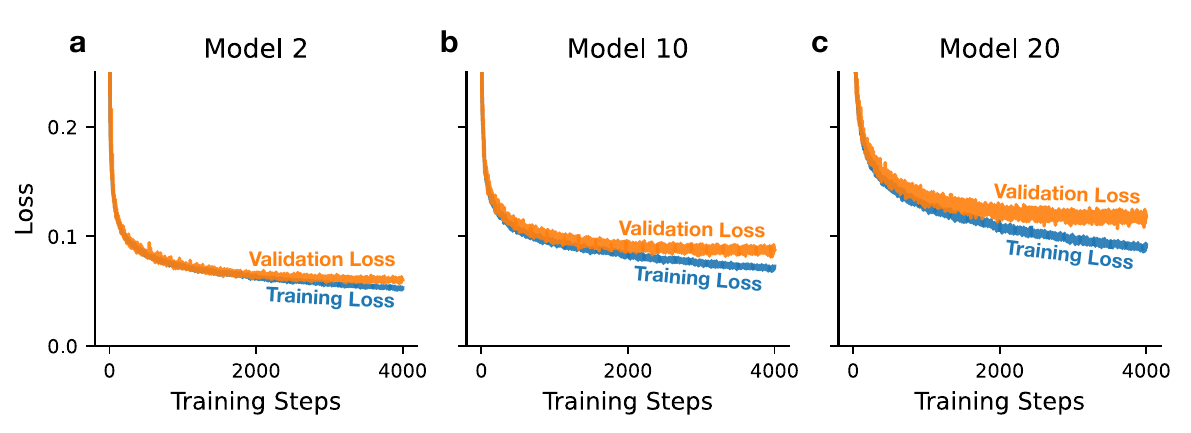}
\caption{Loss curves for the neural network models used in this work. All models exhibit stable and monotonic convergence, and are considered fully trained once the validation loss has plateaued. The observed increase in reconstruction loss with sample thickness is consistent with the growing complexity of thicker specimens. This difference is additionally amplified by the $\approx$ 30\% of data which is simulated off zone. The projected potentials of these samples become increasingly complicated with increasing thickness.}
\label{fig:FigS2}
\end{figure}

\begin{table}[h!]
    \centering
    \begin{tabular}{l c c c}
    \hline
    \textbf{Model} & \textbf{Twist} & \textbf{Substitution} & \textbf{Vacancies} \\
    \hline
    \textbf{Model 2}  & Yes & Yes & Yes \\
    \textbf{Model 10} & No & Yes & Yes \\
    \textbf{Model 20} & No & Yes & Yes \\
    \hline
    \end{tabular}
    \caption{To improve robustness and generalization, the training dataset incorporates several classes of structural disorder commonly found in realistic experimental conditions. In all models, atomic substitution is introduced by randomly replacing 10\% of atoms with elements drawn uniformly from atomic numbers $Z \in \{ 1, 2, \dots, 80 \}$. This broad distribution ensures that the network does not overfit to a narrow compositional space and instead learns features that are invariant to atomic species. In addition, 10\% of atomic sites are randomly removed to simulate vacancies, introducing sparsity and local structural perturbations. 
    In the training data of Model 2, a twist defect is applied along the beam direction. This is implemented as a single rotational discontinuity introduced at a random depth within the sample, producing a relative angular misalignment between two regions of the structure. By including both twisted and untwisted configurations in the training set, the network learns to reconstruct structures in the presence of long-range distortions as well as local disorder.}
    \label{table:supp_defects}
\end{table}

\begin{figure}[h!]
\centering
\includegraphics[width=1.0\textwidth]{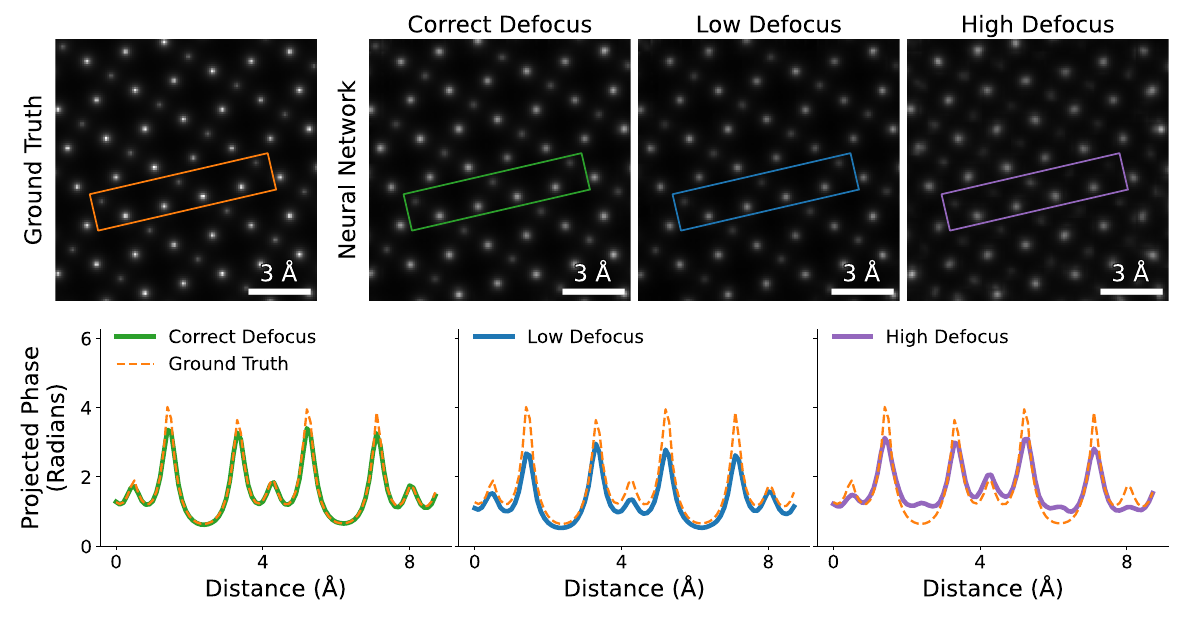}
\caption{Neural network reconstructions with Model 10 on CaPt$_2$ (Materials Project ID mp-842) simulated with different defocus values. Top row: Ground truth potential and neural network reconstructions at the correct defocus (5 nm), low defocus (0 nm) and high defocus (13 nm). Bottom row: Line profiles of the potential averaged across the smaller direction of the highlighted rectangles. The correct defocus (green) tracks the ground truth potential (dashed orange) very well. At low defocus (blue), the neural network systematically underestimates the magnitude of the phase image, while at high defocus (purple), the reconstruction exhibits spatial smearing and spurious peaks in high symmetry locations. While the network is capable of recovering structural features, its qualitative and quantitative accuracy is sensitive to deviations from the training parameter space, leading to systematic artifacts in both amplitude and spatial resolution. }
\label{fig:FigS3}
\end{figure}

\begin{figure}[h!]
\centering
\includegraphics[width=1.0\textwidth]{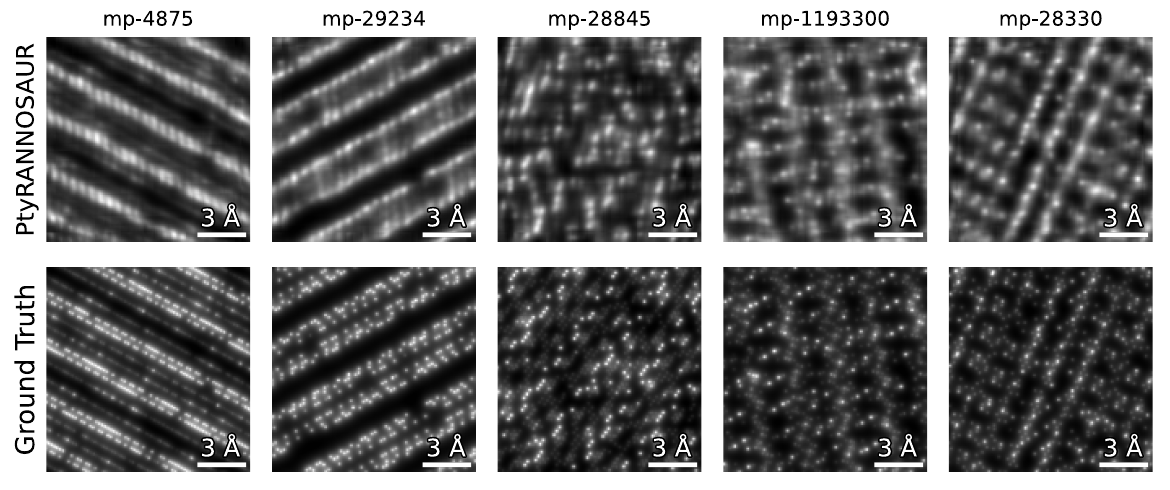}
\caption{Neural network reconstructions on five validation materials using Model 10 with loss of approximately 0.33, corresponding to the 1st (worst) percentile. The materials shown are (from left to right) GdTaO$_4$, InTeI, OsO$_3$F$_2$, ReS$_2$ClO$_9$, and TeAuCl$_7$. While these reconstruction are significantly blurred compared to the ground truth, they still output the general structure, indicating that the network has learned a meaningful representation of the inverse problem that generalizes to complex samples.}
\label{fig:FigS4}
\end{figure} 

\begin{table}[h!]
    \centering
    \begin{tabular}{p{0.5\textwidth} c}
    \hline
    \textbf{Hardware} & \textbf{Approximate Runtime} \\
    \hline
    \textbf{NVIDIA A100 GPU} & 5 seconds \\
    \textbf{NVIDIA RTX PRO 6000 Blackwell} & 5 seconds \\
    \textbf{NVIDIA A10 GPU} & 20 seconds \\
    \textbf{NVIDIA T4 GPU} & 60 seconds \\
    \textbf{GeForce RTX 3050 Ti Laptop GPU} & 60 seconds \\
    \textbf{MacBook Air M4, CPU only} & 15 minutes \\
    \hline
    \end{tabular}
    \caption{Comparison of PtyRANNOSAUR reconstruction runtimes across different hardware platforms. GPU acceleration provides substantial speedups relative to CPU execution, enabling near real-time inference on modern accelerators. This reconstruction time does not include compilation time, which occurs only the first time the code is run, or position correction, which is a 20 second CPU-based post-processing step.}
    \label{table:runtime}
\end{table}

\clearpage

\printbibliography

\end{document}